\documentclass[aps,prb,reprint,showpacs,showkeys,superscriptaddress]{revtex4-1}
\usepackage{multirow}
\usepackage{graphicx}
\usepackage{bm}
\usepackage{dcolumn}
\usepackage{picture}
\usepackage[hidelinks]{hyperref}
\usepackage{natbib}
\usepackage{amsmath}
\usepackage{times}
\usepackage{color}
\usepackage{xcolor}
\newcolumntype{L}[1]{>{\raggedright\let\newline\\\arraybackslash\hspace{0pt}}m{#1}}
\newcolumntype{C}[1]{>{\centering\let\newline\\\arraybackslash\hspace{0pt}}m{#1}}
\newcolumntype{R}[1]{>{\raggedleft\let\newline\\\arraybackslash\hspace{0pt}}m{#1}}
\begin{document}
	
\title{Effect of magnetic and non-magnetic impurities on the spin dimers in the spin 1/2 chains of quantum magnet Sr$_{14}$Cu$_{24}$O$_{41}$}

\author{Rabindranath Bag}\affiliation{Indian Institute of Science Education and Research, Dr. Homi Bhabha Road, Pune, Maharashtra-411008, India}

\author{Koushik Karmakar}\affiliation{Indian Institute of Science Education and Research, Dr. Homi Bhabha Road, Pune, Maharashtra-411008, India}

\author{Sudesh Dhar}\affiliation{Tata Institute of Fundamental Research, Mumbai, Maharashtra, India}

\author{Malvika Tripathi}
\affiliation{UGC DAE Consortium for Scientific Research, Indore 452001, India}

\author{R. J. Choudhary}
\affiliation{UGC DAE Consortium for Scientific Research, Indore 452001, India}

\author{Surjeet Singh}
\email[email:]{surjeet.singh@iiserpune.ac.in}
\affiliation{Indian Institute of Science Education and Research, Dr. Homi Bhabha Road, Pune, Maharashtra-411008, India}\affiliation{Centre for Energy Science, Indian Institute of Science Education and Research, Dr. Homi Bhabha Road, Pune, Maharashtra-411008, India}

\date{\today}
	
	\begin{abstract}
We study the effect of impurities on the two types of spin-dimers in the hybrid chain/ladder spin 1/2 quantum magnet Sr$_{14}$Cu$_{24}$O$_{41}$. Four different impurities were used, namely, the non-magnetic Zn (0.0025 and 0.01 per Cu) and Al (0.0025 and 0.01 per Cu), and magnetic Ni (0.0025 and 0.01 per Cu) and Co (0.01, 0.03, 0.05 and 0.1 per Cu). These impurities were doped in high-quality single-crystals synthesized by the floating zone method. The magnetic susceptibility of pristine Sr$_{14}$Cu$_{24}$O$_{41}$ is analyzed rigorously to confirm that at low temperatures (T $<$ 5 K), the "free" spins in the chains undergo a long-distance dimerization as proposed in a recent study [Sahling et al. Nature Phys., \textbf{11}, 255 (2015)]. The effect of impurity on these dimers is analyzed by measuring the specific heat down to T = 0.06 K. We found that even at the lower impurity concentration, the long-distance dimers are significantly severed, but the quantum entangled spin dimerized state of the chains persists. On the other hand, the other type of spin dimers that forms at relatively higher temperatures via an intervening Zhang-Rice singlet are found to be practically unaffected at the lower impurity concentration; but at 1\% doping, even these are found to be considerably severed. The effect of Co impurity turned out to be most unusual displaying a strongly anisotropic response, and with a dimerization gap that suppresses faster along the chain/ladder direction than perpendicular to it as a function of increasing Co concentration.    
	
 	\end{abstract}
\maketitle
		
	\section{Introduction}
	\label{Intro}
	
Impurities in spin 1/2 chains and ladders can have profound effect on their spin excitations and ground state properties \cite{Alloul2009,Zheludev2013}. Recently, a number of studies have focused on this issue and shown that interesting and rather unexpected properties emerge in the presence of dilute impurities. It has been shown, for instance, that less than 1 \% of Zn or Ni impurities can open a sizable gap in the spin excitation spectrum of spin chain systems SrCuO$_2$ and Sr$_2$CuO$_3$ \cite{SimutisPRL2013,KarmakarPRB2017,KarmakarPRL2017,DalilaPRB2017}; a Co impurity, on the other hand, leaves the excitations gapless but induces a strong Ising-like anisotropy, and a long-range magnetically ordered ground state \cite{KarmakarPRL2017}. In a related 2-leg ladder cuprate SrCu$_2$O$_3$, less than 5 \% of Zn impurity closes the spin-gap and induces a bulk antiferromagnetic ordering of the Cu spins\cite{Azuma1997}. Another exemplary case is that of the spin chain compound CuGeO$_3$, which undergoes a spin-Peierls transition upon cooling below T = 15 K \cite{Hase}. It was shown that impurities in CuGeO$_3$ suppress the Peierls transition and replace it by a long-range antiferromagnetic ground state \cite{Grenier, Uchinokura}.	
	
\begin{figure}[!]
	\centering
	\includegraphics[width=0.45\textwidth]{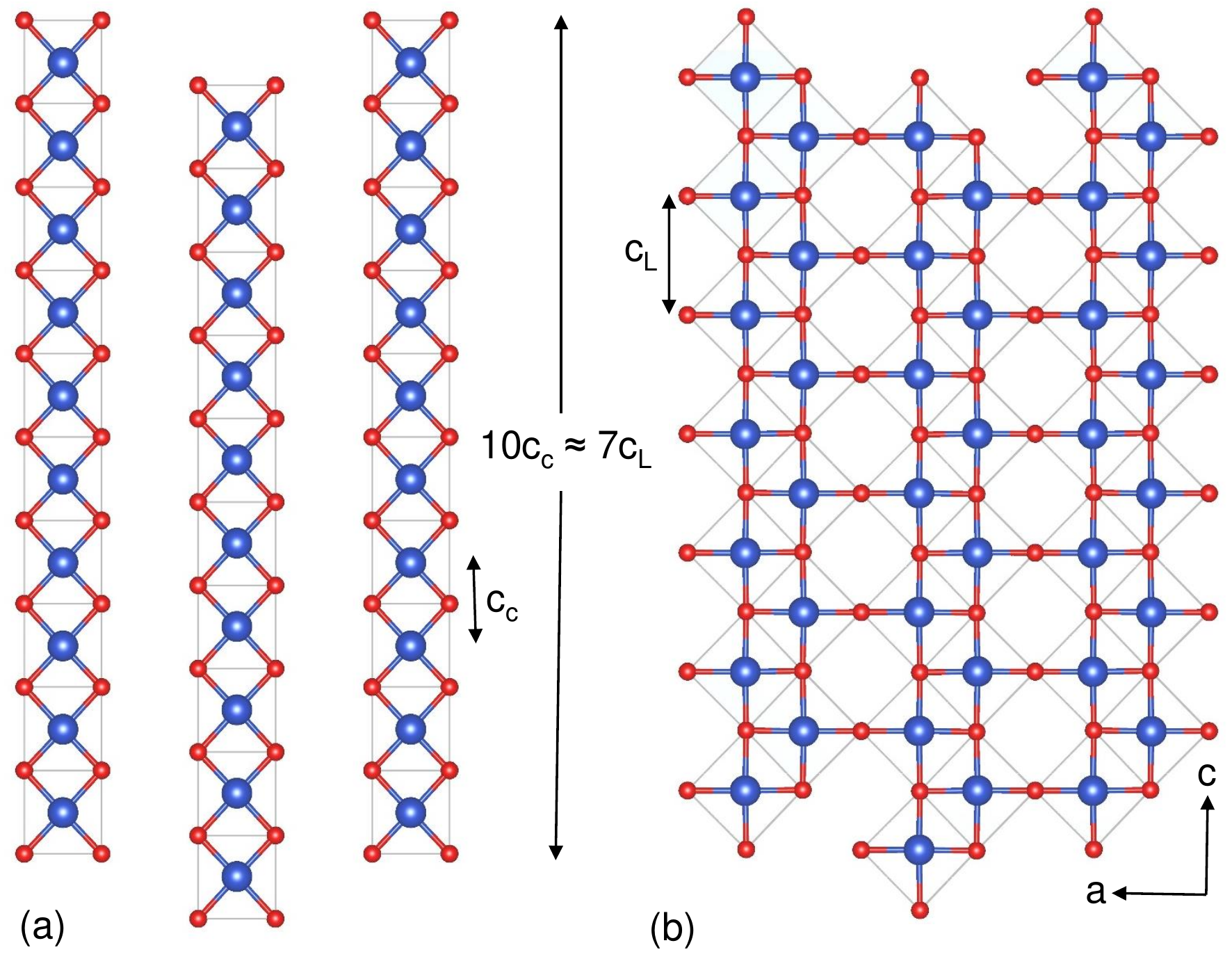}
	\caption {The incommensurate nature of the chain (angle of Cu-O-Cu is 90) and ladder (angle of Cu-O-Cu is 180) sublayers of Sr$_{14}$Cu$_{24}$O$_{41}$ in the ${ac}$ plane.}
	\label{xture}
\end{figure}

A related quantum magnet Sr$_{14}$Cu$_{24}$O$_{41}$ crystallizes with a layered structure comprising an alternating stack of well-isolated ladder and chain planes along the crystallographic \textit{b}-axis. The ladder plane consists of spin 1/2 2-leg Cu$_2$O$_3$ ladders analogous to that in SrCu$_2$O$_3$, and the chain plane consists of 90$^\circ$ spin 1/2 CuO$_2$ chains as in CuGeO$_3$. The chains and ladders run parallel to the c-axis of the orthorhombic unit cell as shown in Fig. \ref{xture} where the chain and ladder sublattices are shown. Interestingly, Sr$_{14}$Cu$_{24}$O$_{41}$ is self-doped with approximately six holes per formula unit. Owing to the presence of holes in the spin 1/2 chain, and a hybrid incommensurate structure (7c$_c$ $\approx$ 10c$_l$, c$_c$ and c$_l$ being the respective lattice parameters of the chain and ladder sublattices along the c-axis), Sr$_{14}$Cu$_{24}$O$_{41}$ exhibits a range of interesting physical properties, including, the charge density wave ordering \cite{VuleticPRL2003}, spin dimerization below 100 K via Zhang-Rice singlets\cite{MatsudaPRB1996}, low-energy pseudo-acoustic sliding density wave modes due to chain/ladder structural incommensurability \cite{ThorsmollePRL2012}, superconductivity under pressure upon Ca doping \cite{UeharaJPSJ1996}, and an exceptionally large magnetic thermal conductivity parallel to the ladders \cite{HessPRB2006}. Recently, it has also been proposed that the spin chains in Sr$_{14}$Cu$_{24}$O$_{41}$ harbor a long-distance quantum entangled ground state comprising dimers between the "free" spins separated by few hundreds of intervening singlets  \cite{SahlingNature2015}. Here, the "free" spins refer to the spins that remain undimerized below 100 K due to a finite fraction of hole transfer from chains to ladders.

In this paper, we investigate the effect of magnetic and non-magnetic impurities by doping at the Cu site with Zn, Ni, Al and Co in Sr$_{14}$Cu$_{24}$O$_{41}$. We first show that at low temperatures, T $\lesssim$ 5 K, the "free" spins in the chains indeed undergo a long-distance dimerization as proposed previously \cite{SahlingNature2015}.  We then study the effect of impurities on the two types of dimers in the chains, namely, the dimers that nucleates at relatively higher temperature (T $\lesssim$ 100 K), and the long-distance dimers that emerge at low-temperatures (T $\lesssim$ 5 K). Using the Ni impurity as a specific test case, we probe the effect of Ni impurity on the long-distance dimers by measuring the specific heat down to a temperature of T = 0.06 K. The effect of Co impurity is investigated up to 10\% of Co doping at the Cu site. The bulk behavior is found to be highly anisotropic even for the lowest Co doping in stark contrast with the nearly isotropic behavior of the pristine crystal. The intradimer excahnge interaction of the high-tenmperature dimers is also shown to exhibit a very anisotropic suppression in the presence of Co doping.     

The manuscript has been organized as follows: in section \ref{EXP} we provide the experimental details. In the results and discussion section \ref{Results}, we first present the susceptibility analysis of the undoped sample in \ref{undoped}. The results of the Zn, Ni and Al impurities are gathered in the section \ref{znnial} under \ref{Zn} for Zn, \ref{Ni} for Ni and \ref{Al} for Al. The results of the Co impurity are presented separately under section \ref{Co}. A summary of all the results and conclusions drawn from this work is gathered under section \ref{sum}.

	\section{Experimental Methods}
	\label{EXP}
Experiments were performed on single-crystals of Sr$_{14}$Cu$_{24}$O$_{41}$, and its doped variants Sr$_{14}$(Cu$_{1-x}$M${_x}$)$_{24}$O$_{41}$ (M = Zn,  x = 0.0025, 0.01; M = Al, x =0.0025,  0.01; M = Ni, x = 0.0025 and 0.01; and M = Co, x = 0.01, 0.03, 0.05 and 0.1) grown using a four-mirror optical floating-zone furnace (Crystal System Corporation, Japan) by the Traveling-Solvent Floating-Zone (TSFZ) method. Polycrystalline rods for the growth experiments were synthesized using the standard solid-state reaction route. The details of growth experiments and structural characterizations of the pristine and Co-doped crystals are previously reported\cite{BagJCG2017}. Single crystals of Zn and Ni doped variants were grown and characterized analogously. The use of high-purity starting precursors (purity $>$ 99.99 \%), along with the use of a crucible free crystal growth technique ensured very high-purity of the grown crystals, free from unintended impurities either from the crucible material or the starting precursors. The high-purity of the crystals is of utmost importance when studying the effect of dilute impurities.

The phase purity of the grown crystals was assessed using powder x-ray diffraction done on crushed single crystal pieces obtained from the grown crystal boule (Bruker D8 advance). For assessing the crystal quality, scanning electron microscopy (Zeiss Ultra Plus) equipped with energy dispersive x-ray (EDX) analysis, and polarized light microscopy (Zeiss Lab A1, AXIO) were done. Details of structural characterizations can be found in Ref. \citenum{BagJCG2017}. Inductively Coupled Plasma - Atomic Emission Spectroscopy (ICP-AES) at the Sophisticated Analytical Instrument Facility (SAIF), Indian Institute of Technology, Bombay is used to determine the actual dopant concentration in the 0.25\% and 1\% Zn, Ni, Al and Co doped crystals. For the Ni and Co doped crystals, the concentration determined using ICP agreed nicely with the nominal value. But in the case of Zn, the actual Zn concentration is found to be 0.004/Cu (for 0.25\%), 0.008/Cu (for 1\%); and for the Al-doped crystal this was around  0.001/Cu (for 0.25\%), 0.0025/Cu (for 1\%), indicating a very limited solid-solubility of Al in the crystals grown using the floating-zone method. Henceforth, the actual concentration is designated as x$_M$, where M can be Zn, Ni, Al or Co.   

Crystals were oriented using the x-ray Laue diffraction technique (Photonics Science, UK).  Magnetic measurements were carried out in the temperature range for 2 to 300 K, and by applying fields up to $\pm$8 Tesla using the vibrating sample magnetometer (VSM) probe in a Physical Property Measurements System (PPMS). Some of the magnetic measurements were also carried out using the SQUID magnetometer in a Magnetic Property Measurement System (MPMS), Quantum Design, USA (maximum field limit $\pm$7 Tesla). Low-temperature specific heat (0.06 K $<$ T $<$ 2 K) was measured at the Tata Institute of Fundamental Research (TIFR) Mumbai in a dilution refrigerator, Quantum Design, USA. X-ray Photoelectron Spectroscopy (XPS) experiments were carried out at the synchrotron INDUS beamline BL-2 AIPES, Indore, India.

\section{Results \& discussion}
\label{Results}
\subsection{Undoped Sr$_{14}$Cu$_{24}$O$_{41}$}
\label{undoped}
\begin{figure}[!]
	\centering
	\includegraphics[width=0.48\textwidth]{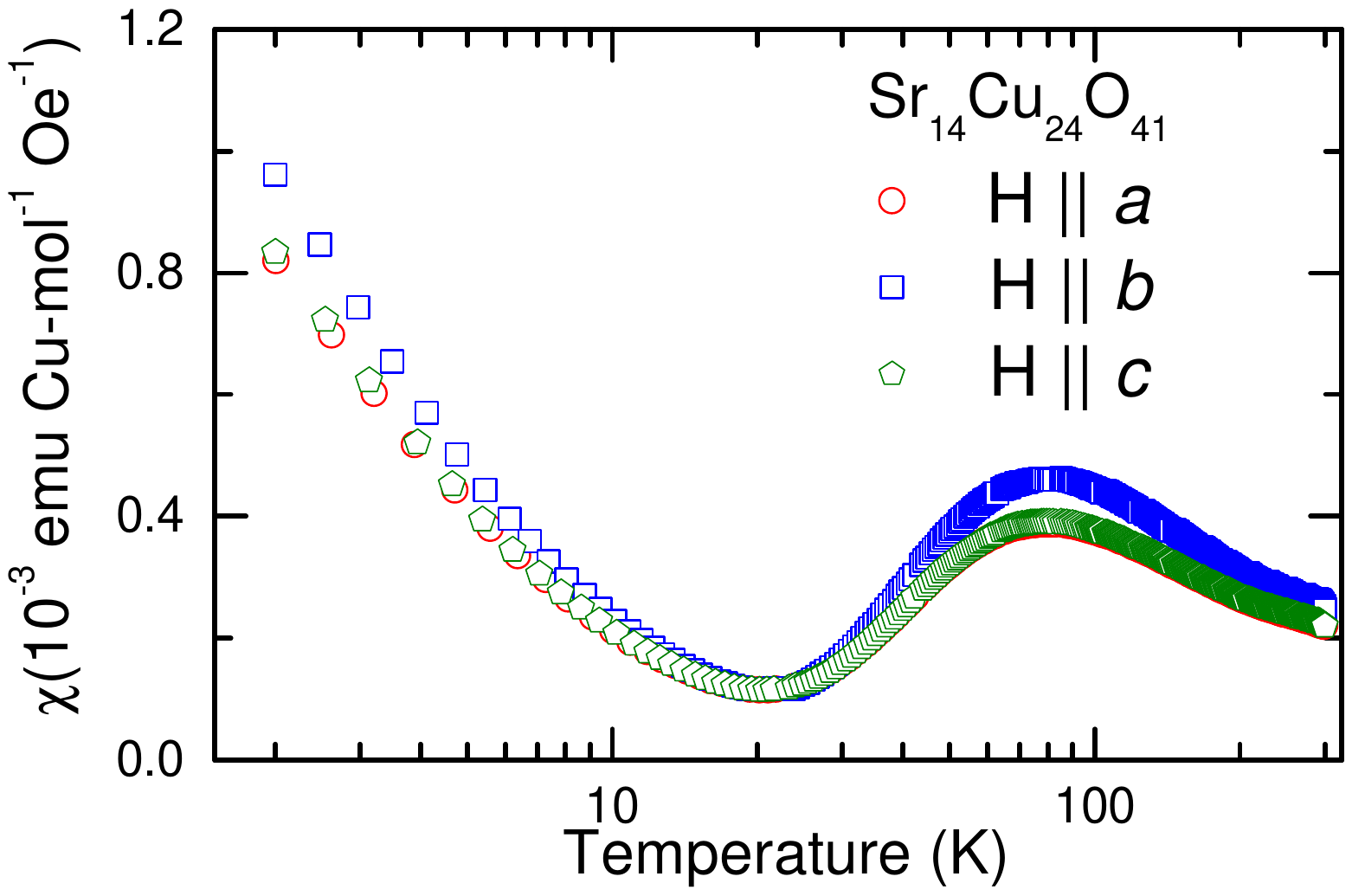}
	\caption{Temperature variation of susceptibility of a pristine Sr$_{14}$Cu$_{24}$O$_{41}$ single crystal along the crystallographic a, b and c-axis} 
	\label{chi_abc}
\end{figure}

The magnetic susceptibility $\chi$(T) of Sr$_{14}$Cu$_{24}$O$_{41}$ along the crystallographic a, b and c-axis is shown in Fig. \ref{chi_abc}. It is evident from these plots that the magnetic behavior of Sr$_{14}$Cu$_{24}$O$_{41}$ is nearly isotropic. The small anisotropy between the in-plane and out-of-plane susceptibility, i.e., $\chi_a = \chi_c \lesssim \chi_b$ (here the subscript represents the crystallographic axis) is due to a slight anisotropy of the Land\'{e} g-factor as reported previously \cite{KlingelerPRB2006}. 

Below room-temperature, $\chi$(T) of Sr$_{14}$Cu$_{24}$O$_{41}$ consists of a broad maximum near T$_{max}$ = 80 K, followed by a sharp upturn below T$_{min}$ = 20 K. Since the ladders in Sr$_{14}$Cu$_{24}$O$_{41}$ have a singlet ground state with a fairly large energy gap ($\Delta_l >$ 500 K) \cite{Tsuji470gapJPSJ,Kumagaispingap500}, the magnetic behavior over the temperature range shown in Fig. \ref{chi_abc} can be attributed mainly to the chain sublattice \cite{KlingelerPRB2006}.

From the previous studies it is known that in a pristine Sr$_{14}$Cu$_{24}$O$_{41}$ crystal majority of the self-doped holes reside over the chain sublattice \cite{Dagotto1999}. The hole in the chain antialign its spin with that of a native Cu$^{2+}$ spin in the chain to form a spin-singlet, which is referred to as the Zhang-Rice (ZR) singlet \cite{ZRsinglet1988}. 

In the chain sublattice there are 10 Cu sites per formula unit (f.u.) \cite{CarterPRL1996}. If one assume all 6 holes to reside over the chain sublattice then the 6 sites in a f.u. will be occupied by the ZR singlets, and remaining 4 by Cu$^{2+}$ spin 1/2 ions. Hence, the ground sate of the chain is proposed to have an arrangement as follows: $\dotsb$ ${\uparrow \bullet \downarrow}$ $\circ$ $\circ$ ${\downarrow \bullet \uparrow}$ $\circ$ $\circ$ $\dotsb$, where the symbols $\bullet$ and $\circ$ represent, respectively, the intra- and interdimer ZR singlets in the chain, and ${\uparrow \bullet \downarrow}$ is a spin dimer in which the spins couple antiferromagnetically via the ZR singlet $\bullet$. Henceforth, we shall refer to the dimer ${\uparrow \bullet \downarrow}$ in the chain as a Zhang-Rice dimer (ZRD).

This configuration of spins and dimers shown above is, however, valid only if all 6 holes reside over the chain sublattice. In a real crystal, due to small fraction of holes transferred to the ladders, the number of holes in the chain sublattice is always slightly less than 6/f.u. (see for example Ref. \citenum{HiroiPRB1996}). As a result, a small but finite fraction of spins in the chain remains undimerized, these are referred to as the "free" spins. 

In Fig. 2, the susceptibility maximum ($\chi_{max}$) near T = 80 K is due to the nucleation of ZRD's, and the upturn below T$_{min}$ = 20 K is typically attributed to the "free" spins. Our data are in fairly good agreement with the previous single crystal reports \cite{MatsudaPhysicaB1998, MatsudaPRB1996, KlingelerPRB2006}. At low-temperatures, the magnitude of $\chi$  exhibits a slight difference: the value of $\chi_c$ near T$_{min}$ = 20 K is reported to be 0.4 $\times$ 10$^{-3} \mu_B$/Cu in Ref. \citenum{KlingelerPRB2006}, which is almost twice the value for our crystal at the same temperature. This difference is most likely related to the purity of starting precursors as discussed later \cite{HiroiPRB1996}.

\begin{figure*}[!]
	\centering
	\includegraphics[width = 0.8 \textwidth]{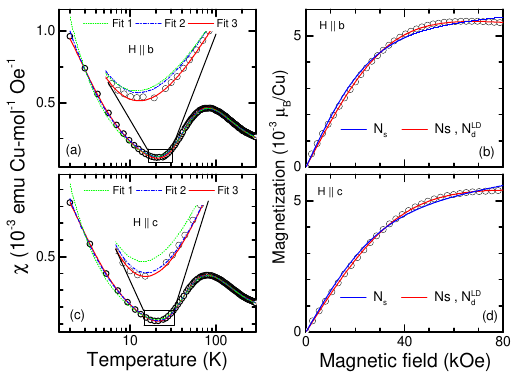}
	\caption{In panels (a) and (c) susceptibility of pristine Sr$_{14}$Cu$_{24}$O$_{41}$ single crystal is shown as a function of temperature along the crystallographic b and c-axis. Fit 1, Fit 2 and Fit 3 refers to the three different fitting models (see text for details). Panels (b) and (d) show isothermal magnetization at T = 2 K as a function of applied magnetic field. Lines are fit to the data (see text for details).} 
	\label{CW_MT_MH}
\end{figure*}

\begin{table*}
	\caption{Values of the various fitting parameters obtained by analyzing the susceptibility of a pristine Sr$_{14}$Cu$_{24}$O$_{41}$ single crystal specimen using the Fit 1 and Fit 2 (see text for details). Corresponding values from the previous reports in the literature are also included}
	\begin{tabular}{L{2.6 cm}    C{2 cm}    C{1.5 cm}    C{1.5 cm}    C{1.2 cm}    C{1.5 cm}    C{2 cm}   C{1.2 cm}    C{1.2 cm}}
		\hline
		\hline
		\vspace{0.1 cm}
		Reports & Specimen type & $\chi_0$ & N$_s$$^\dagger$ & $\theta_P$ & N$_d^{ZR}$ & N$_s$ + 2 N$_d^{ZR}$  &  J$_{ZR}$ & $^\ddagger$g \\
		&  & (10$^{-5}$) & (10$^{-3}$/\textit{Cu}) & (K) & (10$^{-2}$/\textit{Cu}) & (1/\textit{f.u.})  &  (K) &  \\
		\hline
		
		Carter et al.\cite{CarterPRL1996}  &  Polycrystal & - & 22.9 & 0.25 & 6.1 & 3.49 & 139 & 2.17\\
		\hline
		Klingler et al. \cite{KlingelerPRB2006} & H $\parallel$ \emph{c} & 1.0  & 10.0 & - & 7.4 & 3.78 & 134 & 2.04\\
		\hline
		\multirow{3}{2.5cm}{Matsuda et al.\cite{MatsudaPRB1996}} &H $\parallel$ \emph{c} & 0.5 & 15.4 & -1.7 & - & - & 133 & 2.04\\
		& H $\parallel$ \emph{b} & - & 18.3 & -1.7 & - & - & - & 2.26\\
		&	H $\parallel$ \emph{a}& - &15.8 & -1.7 & - & - & - & 2.05\\
		\hline
		Hiroi et al.\cite{HiroiPRB1996} & Polycrystal & 4.4 & 6.6 & 0.2 & 6.7 & 3.4 & 131 & 2.2\\
		\hline
		\multirow{3}{2.6cm}{Present Work (Fit 1)} & H $\parallel$ \emph{c} & 1.9  & 4.7(2) & NU & 7.3(2) & 3.61 & 134(2) & 2.05 \\
		& H $\parallel$ \emph{b} & -2.7 & 4.6(2) & NU & 7.7(2) & 3.79 & 135(2) & 2.25 \\
		& H $\parallel$ \emph{a} & 1.7 & 4.7(2) & NU & 7.4(2) & 3.65 & 134(2) & 2.05 \\
		\hline
		\multirow{3}{2.6cm}{Present Work (Fit 2)}& H $\parallel$ \emph{c} & -2.2 & 6.6(2) & -1 & 7.9(2) & 3.95 & 133(2) & 2.05 \\
		& H $\parallel$ \emph{b} & -3.6 & 6.1(2) & -0.8 & 8.0(2) & 3.98 & 133(2) & 2.25 \\
		& H $\parallel$ \emph{a} & -1.6 & 6.4(2) & -0.9 & 7.9(2) & 3.97 & 134(2) & 2.05 \\
		\hline
		\hline
	\end{tabular}
	\\
	$\chi_0$ is in emu Cu-mol$^{-1}$Oe$^{-1}$; $\ddagger$: g-value is kept fixed, $\dagger$: N$_s$ is obtained from C, NU: parameter not used in the fitting.
	\label{MTreport}
\end{table*}

Until very recently, the role of "free" spins in Sr$_{14}$Cu$_{24}$O$_{41}$ did not attract much attention; and their contribution to the total susceptibility was modeled using either the Curie C/T or the Curie-Weiss law C/(T - $\theta_p$), where C is the Curie constant and $\theta_p$ is the Weiss temperature. Interestingly, Sahling et al. proposed that these "free" spins are quantum entangled and form spin dimers over long-distances (approximately 200 $\AA$ or more) at low temperatures\cite{SahlingNature2015}.  It is, therefore, interesting to investigate further the role of "free" spins.

Here, we first show that the low-temperature $\chi$ of pristine Sr$_{14}$Cu$_{24}$O$_{41}$ crystal cannot be described satisfactorily using either the Curie or the Curie-Weiss law. We then use the model suggested by Sahling et al. to account for the magnetic behavior of the "free" spins, and to get a satisfactory fit to both $\chi(T)$ and M(H) data.    

In previous studies, the magnetic susceptibility of Sr$_{14}$Cu$_{24}$O$_{41}$ at low temperatures has been described using an antiferromagnetic (AFM) spin-dimer model\cite{CarterPRL1996, TakigawaPRB1998, MatsudaPRB1996,MatsudaPhysicaB1998,HiroiPRB1996,KlingelerPRB2006}: 

\begin{equation}
\centering
\chi = \chi_{0} + \chi_C + \chi_{dimer}  
\label{1}
\end{equation}

In this equation, $\chi_0$ represents the temperature independent contribution due to core-diamagnetism and van-Vleck paramagnetism. The second term $\chi_C = C/(T-\theta_p)$ is the Curie-Weiss susceptibility due to the "free" spins. Here, the Curie constant is given by $C= N_{A}N_{S}g^{2}\mu_{B}^{2}S(S+1)/3k_{B}T$; where S is the spin quantum number, N$_s$ is the number of free spins, N$_A$ is the Avogadro number, g is the Land\'e factor, $\mu_B$ is the Bohr magneton, and k$_B$ is the Boltzmann constant. The last term in the above equation represents the dimer susceptibility, $\chi_{dimer}$, given by:
\begin{equation}
\centering
\chi_{dimer} = \frac{2N_{d}^{ZR}N_{A}g^{2}\mu_{B}^{2}}{k_{B}T(3+e^{-\frac{J_{ZR}}{k_{B}T}})}   
\label{dimer}
\end{equation}

where N$^{ZR}_d$ is the number of ZRDs per f.u., and J$_{ZR}$ is the intradimer exchange via the ZR singlet.  

It is possible to include the high temperatures data in the fitting procedure by incorporating an extra term in eq.\ref{1} describing the ladder susceptibility which is given by: $\chi_{ladder} = Ae^{-\Delta_l/k_{B}T}/\sqrt{T}$. Because of the large spin gap $\Delta_l >$ 500 K, which protects the singlet ground state of the ladders, $\chi_{ladder}$ becomes significant only at temperatures above 200 K\cite{Tsuji470gapJPSJ,Kumagaispingap500}.

We fitted our susceptibility data along the three crystallographic orientations using equation \ref{1}, first by fixing $\theta_p = 0$ (this is called the Fit 1); and then by varying $\theta_p$ (this is called the Fit 2). The value of g-factor in these fittings is kept fixed to 2.05 (a-axis and c-axis), and 2.25 (b-axis) based on a previous ESR study\cite{KataevPRB2001}. All other parameters were varied and their best-fit values are listed in Table \ref{MTreport}. The results of similar analysis from previous studies on various polycrystalline and single-crystalline samples are also included in Table \ref{MTreport} for comparison. 

The best-fit value of $\chi_{0}$ in each case is of the order of 10$^{-5}$ emu/Cu-mol, which is very small compared to the measured susceptibility. The value of intradimer coupling (J$_{ZR}$) in our sample is inferred to be 134 $\pm$ 2 K which is in a fairly good agreement with previous single crystal reports. As for the number of dimers per formula unit (N$_d^{ZR}$), we found it to be around 0.08/Cu for the Fit 2, and to vary between 0.07/Cu and 0.08/Cu for Fit 1 depending on the crystal orientation. These values are slightly higher than those previously reported but do not overshoot the upper bound of N$_d^{ZR}$ = 2/f.u. = 0.083/Cu. Finally, N$_s$ for our crystal is smaller than the values previously reported, which in the polycrystalline sample \cite{CarterPRL1996} is almost 4 times the values found here for high-purity single crystals. The sum N$_s$ + 2N$_d^{ZR}$, which should be 4 in the ideal case when all the holes are localized on the chain sublattcie, varies from 3.61/f.u. to 3.79/f.u. for Fit 1 and it is 3.98/f.u. for Fit 2.

The quality of Fit 1 and Fit 2 can be judged from the fitted curves in Fig. \ref{CW_MT_MH} (left panels). Since $\chi_a$ = $\chi_c$, data along the a-axis is not shown. Quite generally, we found that both the fits reproduce the qualitative behavior of $\chi(T)$ reasonably well. However, a closer inspection reveals the presence of temperature regions where these fits deviate from the measured data. For instance, Fit 1 is clearly not satisfactory below about T = 5 K, and it also deviates from the measured data near T = 20 K, where $\chi$ takes its minimum value. On the other hand, while the Fit 2 is more acceptable at low temperatures, it deviates from the measured data marginally around $\chi_{min}$, as shown in the insets of  Fig. \ref{CW_MT_MH} (left panels). Moreover, in the absence of any magnetic ordering down to at least 60 mK \cite{SahlingNature2015}, $\theta$$_p \sim$ 1 to 2 K obtained in Fit 2 is hard to explain.

According to Sahling et al. \cite{SahlingNature2015}, at low-temperatures, the "free" spins are quantum entangled; i.e., they form spin dimers over large-distances via an AFM interaction mediated via the intervening Cu-sites. We shall denote the exchange between these dimers by J$_{LD}$, where the subscript LD refers to the long-distances between the pairing spins. The long distance dimers are designated as $\uparrow \dots \downarrow$, where $\dots$ represent the intervening sites. We, therefore, fitted the measured $\chi$ by including an extra dimer term in eq.\ref{1} to account for the long-distance dimer (LDD), and by setting $\theta_p$ to zero, which is not required in this model. The modified equation for $\chi$ is now given by:

\begin{equation}
\chi = \chi_{0} + \frac{C}{T} + \frac{2N_{A}g^{2}\mu_{B}^{2}}{k_BT}\sum_{i}^{} \frac{N_{d_i}}{(3+e^{{-\frac{J_i}{k_{B}T}}})}
\label{2}
\end{equation}

Here, N$_{d1}$ = N$_d^{ZR}$ and J$_{1}$ = J$_{ZR}$, as before corresponds to the Zhang-Rice dimer (ZRD). N$_{d2}$ and J$_2$ are analogous parameters for the LDD, where N$_{d2}$ = N$_{d}^{LD}$ and J$_{2}$ = J$_{LD}$. The fit to the measured $\chi$ using equation \ref{2} is called the Fit 3. 

\begin{table*}
	\centering
	\caption{Fitting parameters obtained from an analysis of the susceptibility of undoped Sr$_{14}$Cu$_{24}$O$_{41}$ using the Fit 3 (see text for deatils)}
	\begin{tabular}{L{1.3 cm}C{1.5 cm}C{1.5 cm}C{1.5 cm}C{1.5 cm}C{1.5 cm}C{1.5 cm}C{1.5 cm}}
		\hline
		\hline
		\vspace{0.1 cm}
		&  $\chi_0$ & $^\diamond$N$_s$  & N$_d^{ZR}$ & J$_{ZR}$ &  N$_{d}^{LD}$  &  J$_{LD}$ & $^\dagger$g  \\
		&  (10$^{-5}$)  & (10$^{-3}$/\textit{Cu}) & (10$^{-2}$/\textit{Cu}) & (K) &  (10$^{-2}$/Cu)  &  (K) &\\
		\hline
		H $\parallel$ \emph{c} & -2.4 & 2.8(2) & 7.9(2)& 132(2) & 0.17(5) & 4(1) & 2.05 \\
		H $\parallel$ \emph{b} & -5.4 & 2.5(2) & 8.0(2)& 133(2)& 0.17(5) & 4(1) & 2.25 \\
		H $\parallel$ \emph{a} & -2.7 & 2.5(2) & 8.0(2) & 132(2) & 0.17(5) & 4(1) & 2.05\\
		\hline
		\hline
	\end{tabular}
	
	$\chi_0$ is in emu Cu-mol$^{-1}$Oe$^{-1}$ unit. $\dagger$ g-value is kept fixed during the fitting, $\diamond$ N$_s$ is calculated from the Curie constant (C)
	\label{t2}
\end{table*}

The results of Fit 3 are also included in Fig. \ref{CW_MT_MH} (left panels). We found that the Fit 3 represents the experimental data more accurately than Fit 1 or Fit 2. The fitting parameters in Fit 3 are listed in Table \ref{t3}. The values of J$_{ZR}$ and N$_d^{ZR}$ are nearly the same as obtained earlier using the Fit 2. N$_s$ has slightly decreased, which is understandable since some of the free spins have dimerized at low-temperatures. The value of J$_{LD}$ is $\approx$ 4 $\pm$ 2 K and that of N$_d^{LD}$ $\sim$ 0.0017/Cu. From these numbers we can estimate the quantity N$_S$ + 2N$_d^{ZR}$ + 2N$_{d}^{LD}$, which gives the effective number of spins 1/2 in the chain sublattice as 0.162/Cu $=$ 3.88/f.u. $\approx$ 4/f.u. This value being not exactly equal to 4 suggests that a small fraction of holes resides over the ladder sublattice.

To show further the poor applicability of Curie- or Curie Weiss law alone near T = 2 K, we first analyzed the T = 2 K isothermal magnetization of Sr$_{14}$Cu$_{24}$O$_{41}$ using the spin 1/2 Brillouin function with N$_s$ as a free parameter as in eq. \ref{bri}:

\begin{equation}
M(H) =\chi_{0}H+N_{s}N_{A}g\mu_{B}B_{\frac{1}{2}}(y)
\label{bri}
\end{equation}

where, $H$ is the applied field and S = 1/2; $y = g\mu_{B}HS/k_{B}T$ and $B_{\frac{1}{2}}(y)$ is the spin 1/2 Brillouin function. In this equation, a small contribution from $\chi_0$ is also included in the first term. Since the contribution of ZRDs will be negligible at T = 2 K, eq. \ref{bri} should fit the 2 K isotherm if the "free" spins are not dimerized or magnetically ordered. 

The results are shown in Fig. \ref{CW_MT_MH}(right panels). The fitted curve (blue line in panels b and d of Fig. \ref{CW_MT_MH}) deviates from the measured data over the whole range for both field orientations. The deviations are not very large but they are significant enough to suggest that eq. \ref{bri} is not sufficient to fit the M(H) data at 2 K satisfactorily, corroborating the need for considering an additional term to account for the weak AFM interaction between the "free" spins.

When a term accounting for LDDs is incorporated in eq. \ref{bri}, an excellent fit to the data is obtained for both field orientations. This is shown by the red line in panels b and d of Fig. \ref{CW_MT_MH}. At temperatures higher than 10 K, the ZRDs gain importance while the contribution of LDDs become very small; M(H) fitting for higher temperatures is given separately in the supplementary information.

\subsection{Zn, Ni and Al impurities}
\label{znnial}
We now present the effect of doping Zn, Ni and Al impurities on the magnetic behavior of Sr$_{14}$Cu$_{24}$O$_{41}$. $\chi(T)$ of 0.25\% and 1\% Zn, Ni and Al doped crystals is shown in Fig. \ref{ZnAlNi}. The data for pristine sample are also included for comparison. For brevity, only the data gathered by applying magnetic field parallel to the chains/ladder direction (i.e., H $||$ c) are shown. The results for other orientations are similar and they are shown separately in the supplementary information. Since in Co case the effect of impurity on the magnetic behavior is far more pronounced, we deal with it in a separate section.

For the convenience of our reader, we recollect here the various symbols that we will be using extensively in the subsequent sections: 

$\uparrow$ or $\downarrow$ for a Cu$^{2+}$ spin 1/2; $\Uparrow$ for a Ni$^{2+}$ spin 1; $\circ$ is for an inetrdimer ZR singlet; $\bullet$ is for an intradimer ZR singlet; $\uparrow \bullet \downarrow$ represents a ZR dimer (ZRD), and $\uparrow \dots \downarrow $ is for a long-distance dimer (LDD).  

 \begin{table*}
	\centering
	\caption{Fitting parameters obtained from an analysis of the c-axis susceptibility for Zn, Ni and Al impurities (see text for details)}
	\begin{tabular}{C{1.2 cm} C{1.2 cm}  C{1.7 cm}  C{1.7 cm} C{1.7 cm} C{1.5 cm}  C{1.5 cm}  C{1.7 cm}  C{1.5 cm}  C{1.7 cm}}
		\hline
		\hline
		\
		Dopant & x &    Fitting-range    & $\chi_0$ & $\theta_p$ & C & Spin  & $^\diamond$N$_s$ & N$_d^{ZR}$  & J$_{ZR}$ \\
		
		(M) & (\%) & (K) & (10$^{-5}$) & (K) & (10$^{-2}$/\textit{Cu}) & (S)  & (10$^{-3}$/\textit{Cu})  & (10$^{-2}$/\textit{Cu}) & (K)  \\
		
		\hline
		
		Undoped & 0 & 10 - 200 & -1.9 & NU & 0.22(2) & 1/2 & 5.9(2) & 7.8(2) & 132(2) \\
		Zn & 0.25 & 10 - 200 & -0.9 & NU & 0.18(2) & 1/2 & 5.1(2) & 7.8(2)  & 130(2) \\
		Ni & 0.25 & 10 - 200 & -1.6 & NU & 0.44(2) & 1 & 4.4(2) & 7.8(2) & 131(2)  \\
		Al & 0.25 & 10 - 200 & -1.5 & NU & 0.17(2) & 1/2 & 4.6(2) & 7.8(2) & 128(2) \\
		\hline
		Zn & 1 & 10 - 200 & 1.0 & NU & 0.45(2) & 1/2 & 12.1(4) & 6.9(4) & 125(2) \\
		Ni & 1 & 10 - 200 & 0.8 & NU & 1.14(2) & 1 & 11.4(4) & 7.1(4) & 128(2) \\
		Al & 1 & 10 - 200 & -2.2 & NU & 0.29(2) & 1/2 & 7.9(4) & 7.5(4) & 127(2) \\
		\hline
		\hline
	\end{tabular}
	
	The unit of $\chi_0$ is emu (Cu + M )-mol$^{-1}$ Oe$^{-1}$, $^\diamond$ N$_s$ is obtained from C; g-value is 2.05. NU: parameter is not used. 
	\label{t3}
\end{table*}

\subsubsection{Zn impurity}
\label{Zn}
\begin{figure*}[hbtp]
	\centering
	\includegraphics[width = 0.90 \textwidth]{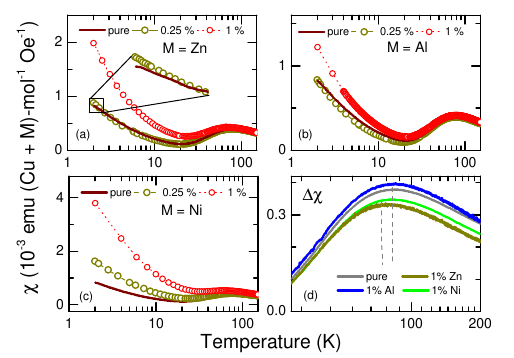}
	\caption{Temperature variation of susceptibility of (a) Zn, (b) Al and (c) Ni doped Sr$_{14}$Cu$_{24}$O$_{41}$ single crystals along the crystallographic c-axis. In (d) the dimer contribution is shown for each dopant. In each panel data for the pristine crystal is shown for comparison. } 
	\label{ZnAlNi}
\end{figure*}

 We first consider the case of a Zn impurity which is expected to be in the Zn$^{2+}$ (3d$^{10}$) state. Due to a completely filled d-shell, a Zn$^{2+}$ ion is non-magnetic. As shown in Fig. \ref{ZnAlNi}, $\chi(T)$ of 0.25\% Zn doped crystal (Zn0.25) overlaps that of the undoped crystal except below T = 2.5 K where it begins to exceed marginally, as shown in the inset. This is an interesting result which suggests that at low concentrations, the Zn ions substitute $\circ$ in the chain because in any other event (i.e., Zn substituting $\uparrow$ or $\bullet$), "free" spins will be liberated which will enhance the Curie-tail.

When the doping concentration is raised to 1\% (Zn1), the low-temperature susceptibility enhances by nearly 200 \%. This strong enhancement suggests that at this higher doping concentration, Zn substitutes not only $\circ$ in the chain but also $\bullet$ and/or $\uparrow$ (or $\downarrow$) in $\uparrow \bullet \downarrow$. In any of these events free spins will be liberated. However, which one of these is more favored can be determined by analyzing $\chi(T)$ quantitatively.  

We analyzed $\chi$(T) of Zn0.25 and Zn1 crystals in the temperature range 10 K to 200 K by using eq. \ref{2} but without considering the LDD term, whose contribution above 10 K is expected to be negligible in any case. We excluded the temperature range below 10 K since the precise nature of the magnetic ground state in the presence of Zn is not known.  

To test the validity of this procedure, we first fitted $\chi(T)$ of the undoped sample in this temperature range. This fitting procedure also resulted in the same values of N$_d^{ZR}$ and J$_{ZR}$ as obtained in the previous section (see Tables \ref{t2} and \ref{t3}). The value of N$_s$ did increase slightly but this is not unexpected since the spins in the dimer $\uparrow \dots \downarrow$ also contribute to N$_s$ in this temperature range.  

For Zn0.25, the parameters N$^{ZR}_d$, J$_{ZR}$ and N$_s$ are practically the same as for the undoped crystal, which is to be expected as the two data sets overlap in this temperature range. In the Zn1 crystal, N$_d^{ZR}$ has decreased from 0.078/Cu for the undoped crystal to 0.069/Cu. Simultaneously, J$_{ZR}$ decreased from 132 to 125 K. A significant decrease in the values of these parameters provides a direct evidence that on increasing the doping level to 1\%, the Zn impurity tends to break the dimers $\uparrow \bullet \downarrow$ in the chains.

We note that for Zn1, the value of N$_s$ has increased to 0.012/Cu. If we take into account the value of N$_s$ of the undoped crystal, the effective increase in N$_s$ ($\Delta$N$_s$) due to Zn-doping in Zn1 is $\sim$0.006/Cu. Similarly, $\Delta N_d^{ZR}$ is $\sim$0.009/Cu. These values are in good accord with the actual concentration of Zn in the crystal as determined using the ICP technique.   

\subsubsection{Ni impurity}
\label{Ni}
We now consider the case of a Ni$^{2+}$ (3d$^{8}$) impurity which can either be in a spin 0 (low-spin) or a spin 1 (high-spin) state. To determine the spin state of Ni, we estimated the excess susceptibility ($\Delta\chi$) due to Ni impurity by subtracting $\chi$ per Cu of 1\% Ni doped crystal (Ni1) from that of the undoped crystal. The excess susceptibility, when plotted as $\Delta\chi^{-1}$ versus T, showed a fairly good linearity both above and below 60 $\pm$ 10 K, and with a mild curvature around this temperature, which is reminiscent of the $\chi_{max}$. From the Curie-Weiss analysis of data above T = 150 K, we estimated the spin-state of Ni to be S = 1.  

$\chi$ of Ni0.25 and Ni1 is analyzed along the same line as done previously for the Zn doped crystals. $\chi$(T) of Ni0.25, shown in Fig. \ref{ZnAlNi}, exhibits an enhanced Curie-tail at low-temperatures. In this case, fitting $\chi$ in the range T = 10 K to 200 K resulted in a good-fit with practically unchanged values of N$_d^{ZR}$ and J$_{ZR}$, but with Curie constant increased from 0.22 (undoped) to 0.45 emu (Cu + M)-mol$^{-1}$Oe$^{-1}$K. Taking into account that the "free" spin density in the undoped sample is 0.0022/Cu, the net increase in the "free" spin density ($\Delta N_s$) due to Ni doping comes out to be 0.0023/Cu, which agrees fairly nicely with the Ni concentration ($x_{Ni}$) in the crystal (0.0025/Cu). 

This simple analysis suggests that Ni substitutes $\circ$ in the chains akin to the Zn case discussed above.    

In Ni1, $\Delta$N$_d^{ZR}$ is 0.007/Cu, and  $\Delta$J$_{ZR}$ is about 4 K. From the decrease in values of these parameters, we can infer that at higher Ni concentrations the Ni impurity breaks the ZR dimers $\uparrow \bullet \downarrow$ analogous to the case of Zn. 

A Ni impurity can break a dimer either by substituting $\uparrow$ (or $\downarrow$) in the dimer $\uparrow \bullet \downarrow$ as $\Uparrow \bullet \downarrow$ or $\uparrow \bullet \Downarrow$, where $\Uparrow$ represents the Ni spin. The other possibility is that it substitutes directly at the $\bullet$ site. In order to gain some insight into what is preferred in the Ni case we proceed as follows: 

We know that for Ni1 $\Delta$N$_d^{ZR}$ = $\sim$ 0.007/Cu (see Table \ref{t3}), i.e., due to Ni doping the number of dimers have reduced. To began with, we presume that the probability of $\Uparrow$ substituting a $\bullet$ is relatively weak, i.e., the dimers break with the substitution of a $\Uparrow$ either at $\uparrow$ or $\downarrow$ in $\uparrow \bullet \downarrow$. Whether this presumption is correct or not will become clear at the end depending on what we get. 

After a dimer has been broken, we assume $\uparrow$ and $\Uparrow$ to be "free" to align with the field (i.e., the interaction between $\uparrow$ and $\Uparrow$, mediated via the ZR singlet is weak), and hence both spins will contribute independently to the Curie-tail. Their respective contributions can be calculated using the expression: $C= \eta N_{A}g^{2}\mu_{B}^{2}S(S+1)/3k_{B}T$, where S = 1 for $\Uparrow$ and 1/2 for $\uparrow$, and $\eta$ is 0.007/Cu (= $\Delta$N$_d^{ZR}$) for both. Form this calculation we obtained the Curie constant for $\uparrow$ (C$_{\uparrow}$) and $\Uparrow$ (C$_{\Uparrow}$) as 0.0026 and 0.007 emu mol$^{-1}$ Oe$^{-1}$ K, respectively. 

Above we accounted for 0.007/Cu of Ni ions breaking the dimers. Since $x_{Ni}$ is 0.01/Cu, we still need to account for the remaining 0.003/Cu of Ni ions. From the previous analysis of Ni0.25 (or Zn0.25), we expect these Ni ions to substitute at $\circ$ site in the chains. The Curie constant due to these Ni ions (C$_{\Uparrow}^\circ$) can be calculated as above by substituting $\eta$ = 0.003/Cu in the expression for C. We get, C$_{\Uparrow}^\circ$ = 0.003 emu Cu-mol$^{-1}$Oe$^{-1}$K. 

Hence, the net Curie constant (C) due to Ni impurity in Ni1 will be C$_{\uparrow}$ + C$_{\Uparrow}$ + C$_{\Uparrow}^\circ$ = 0.0126 emu Cu-mol$^{-1}$Oe$^{-1}$K. The experimental value of this quantity can be obtained by subtracting from the Curie constant of Ni1 the Curie constant of the undoped crystal, as given in Table \ref{t3}. This is done to take into account the small number of "free" spins that are expected to be there even in the absence of Ni (as in the undoped crystal). Doing so gives $\Delta$C$_{exp} =$ 0.009 emu Cu-mol$^{-1}$ Oe$^{-1}$ K, which is comparable to the calculated value of 0.0126/Cu.      

Note that substitution of $\Uparrow$ for $\bullet$ in the dimer will further enhance the calculated Curie constant, making the disagreement between the calculated and experimental $\Delta$C worse, which supports the presumption made at the beginning concerning the site preference of Ni. 

From these reasonably credible analyses one can draw the following conclusions: (i) in terms of their doping habits, both, Zn and Ni impurities behave analogously, (ii) at low-doping concentration, at least up to 0.0025/Cu, both, Zn and Ni tend to replace the interdimer ZR singlet in the chain, (iii) at higher concentration, dimers $\uparrow \bullet \downarrow$ are broken, (iv) probability for substituting $\bullet$ in $\uparrow \bullet \downarrow$ is found to be smallest, which makes sense because there are only 2 sites in a f.u. occupied by $\bullet$.       
 
\begin{figure}[!]
	\centering
	\includegraphics[width=0.5\textwidth]{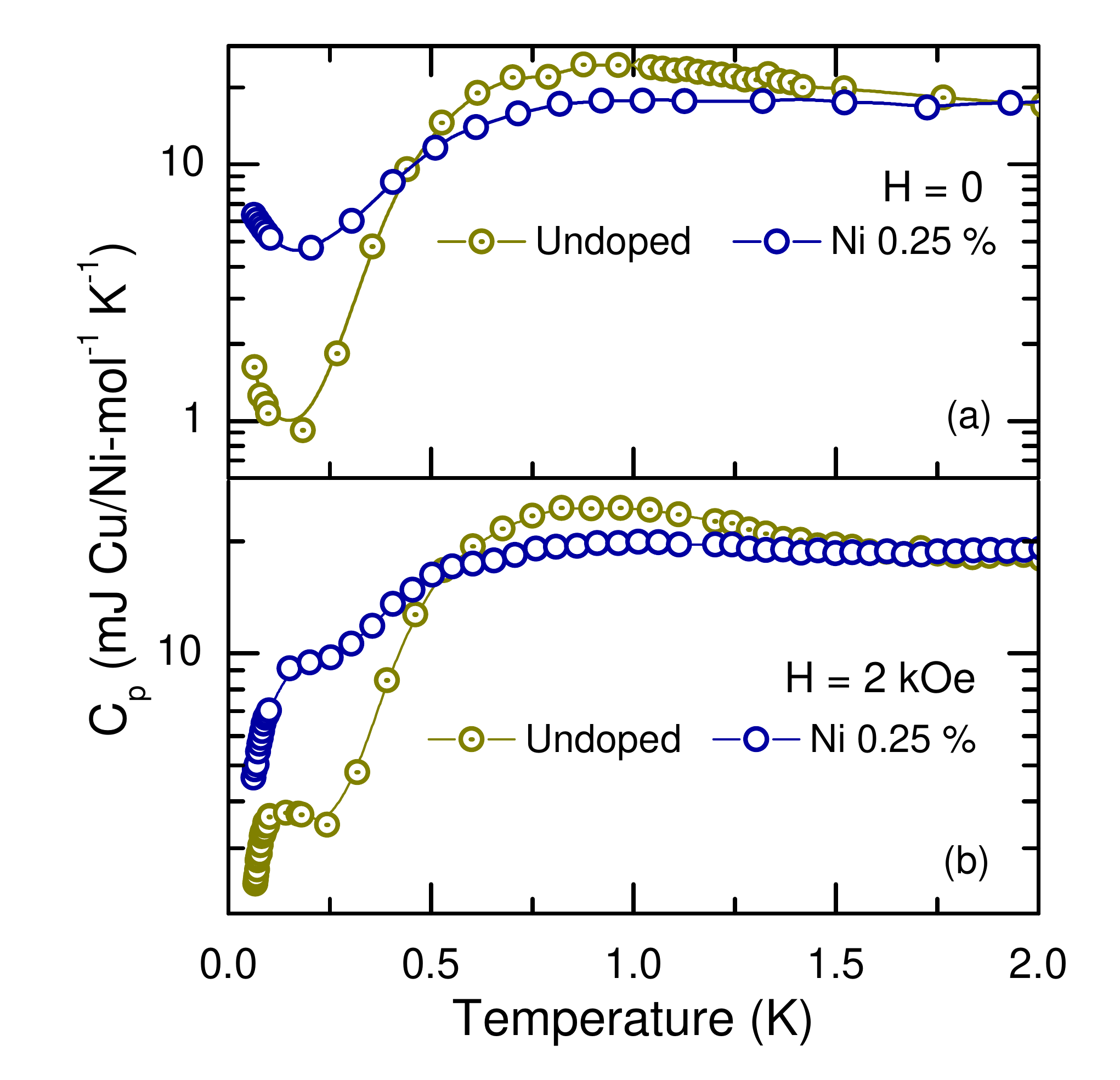}
	\caption{Temperature variation of specific heat (C$_p$) of pristine and 0.25\% Ni doped Sr$_{14}$Cu$_{24}$O$_{41}$ single crystals: (a) under zero-field condition (H = 0), (b) under H = 2 kOe. Lines are drawn as a guide to the eye.} 
	\label{spheat}
\end{figure}

\textit{Specific heat of Ni0.025}: So far we concentrated on the effect of impurities on the ZRDs $\uparrow \bullet \downarrow$, but how their presence in the chains affects the LDDs $\uparrow \dots \downarrow$ remains unevaluated. To understand this, we studied the specific heat of the undoped and Ni 0.25\% doped crystal below T = 2 K. The results of the specific heat (C$_p$) measurements down to a temperature of 0.06 K are shown in Fig. \ref{spheat}. 

We note that near T = 2 K, the value of C$_p$ of the pristine Sr$_{14}$Cu$_{24}$O$_{41}$ is about 25 mJ Cu-mol$^{-1}$K$^{-1}$, which is higher than a value of 10 mJ Cu-mol$^{-1}$K$^{-1}$ previously reported for the spin chain compound Sr${_2}$CuO${_3}$ \cite{KarmakarPRB2017}. The higher value of C$_p$ in Sr$_{14}$Cu$_{24}$O$_{41}$ is due to the presence of "free" spins that undergo long-distance dimerization at low temperatures. This is manifested  in the low-temperature specific heat in the form of a broad maximum centered around T = 0.9 K. Another noteworthy feature in the low-temperature specific heat is the presence of an upturn below T = 0.15 K. This feature is probably the high-temperature tail of the Schottky anomaly that peaks below T = 0.06 K. The Schottky anomaly arises from a small number of "free "spins that have not dimerized due to very large separations; the presence of a weak magnetic field in the cryostat tend to align these spins giving rise to a Schottky anomaly at very low-temperatures\cite{SahlingNature2015}. This interpretation gains more credibility once a small magnetic field is turned on. As shown in panel (b), under a small applied field of 2 kOe, the Schottky anomaly shifts considerably to higher temperatures, and is now peaked around T = 0.2 K. On the other hand, the applied field has no discernible effect on the position of 0.9 K anomaly associated with the long-distance dimers, which is not surprising given that the J$_{LD}$ (4 $\pm$ 2 K) is too large compared to the magnitude of applied magnetic field. 

A quantitative analysis of the low-temperature C$_p$ can be carried out following Sahling et al. \cite{SahlingNature2015}. They approximated the measured specific heat of an undoped Sr$_{14}$Cu$_{24}$O$_{41}$ by expressing it as C$_p$ = C$_{ph}$ + C$_{dimer}$ + C$_{Sch}$, where C$_{ph}$ represents the phononic contribution which can be approximated as $\beta$T$^3$; C$_{dimer}$ is the dimer term which can be expressed as: A($\Delta$/k$_B $T)$^\frac{3}{2}$exp(-$\Delta$/{k$_B$T}), and C$_{Sch}$ is the Schottky contribution\cite{Gopal} of the "free" spins. Here, the coefficient A in the dimer term is simply N$_d^{LD}$ times the gas constant (R), and $\Delta$ corresponds to the exchange J$_{LD}$ (supplementary information).

Let us now turn our attention to how the T = 0.9 K anomaly associated with the long-distance dimers is affected due to a small amount of Ni doping. In Fig. \ref{spheat}, we notice that C$_p$ of the Ni doped crystal exhibits the same qualitative features as found for the undoped crystal, but with the exception that the 0.9 K anomaly has been suppressed, and at the same time C$_p$ at temperatures less than 0.5 K has been enhanced. The enhancement below 0.5 K can be attributed to the Ni spins and the new "free" spins that may have released due to a severing of the long-distance dimers upon Ni doping.  The low-temperature C$_p$ in this case is evidently more complex due to the presence of Ni spins (spin 1) in the chain. Since we do not know the exact ground state in the presence of Ni spins, we defer a detailed investigation of C$_p$ below 0.5 K for a future study and focus here on the suppression of the T = 0.9 K anomaly associated with the long-distance dimers, which is what we had started with.

From the magnitude of decreases in the height of T = 0.9 K anomaly, the suppression of long-distance dimers can be estimated to be $\approx$ 28\%. A slightly more refined value of 21\% is obtained by fitting C$_p$ as discussed above (see supplementary information for details). This suppression suggests a severing of the long-distance dimers. Given that the impurity concentration is merely 0.0025/Cu, which corresponds to nearly 1 Ni impurity for every 400 Cu sites in the chain sublattice, the severing appears to be quite significant for the dilute amount of impurity incorporated. However, the large separation (around 100 Cu sites on an average\cite{SahlingNature2015}) between the spins in the dimer $\uparrow \dots \downarrow$ implies that there is a fair probability of finding a Ni impurity in approximately one out of four randomly picked dimers in the chain, which is in agreement with the magnitude of suppression observed experimentally.

These considerations suggests that only those long-distance dimers that enclose a Ni impurity between the pairing spins as $\uparrow \dots \Uparrow \dots \downarrow$ are severed; and the dimers without a Ni impurity ($\uparrow \dots \downarrow$) are not affected by their presence outside. What we can conclude from here is that while the dimer $\uparrow \dots \downarrow$ itself is sensitive to the presence of a Ni impurity, the long-distance quantum entangled ground state of the spin chains in Sr$_{14}$Cu$_{24}$O$_{41}$ appears to be quite robust to these impurities.         

\subsubsection{Al impurity}
\label{Al}
 The Al impurity is expected to be in a +3 (spin 0) state. As shown in Fig. \ref{ZnAlNi}, in Al0.25, $\chi(T)$ is nearly unchanged compared to the undoped sample. When the doping concentration is increased to 1\%, the low-temperature enhancement is observed but it is much smaller than in the case of Zn. Since both Zn and Al are non-magnetic, the difference between these impurities could be due to a lower solubility limit of Al in the crystal as revealed by the ICP technique. In the Al1 crystal, the actual Al concentration (x$_{Al}$) is close to 0.25\%. 
 
Using the same analysis for treating $\chi(T)$ of crsytal Al1 as done previously for the Zn and Ni doped crystals, we found that the increment $\Delta N_s$ upon Al doping to be $\sim$0.002/Cu, which is consistent with the value of x$_{Al}$. Changes in N$_d^{ZR}$ and J$_{ZR}$ upon Al doping are found to be marginal, which is also reflected in the fact that $\chi_{max}$ shifts only slightly upon Al-doping (panel d). Since Al impurity increases the Curie-tail without breaking the dimers (N$_d^{ZR}$ and J$_{ZR}$ are unchanged), one can tentatively conclude that unlike the divalent impurities, Zn and Ni, the tripositive Al impurity has a tendency to dope the ladders rather than the chains.   
 
\subsection{Co impurity}
\label{Co}
\subsubsection{Susceptibility}
The strongest effect of doping on the magnetic behavior is observed when Sr$_{14}$Cu$_{24}$O$_{41}$ is doped with a Co impurity. As shown below, with merely 1\% of Co doping, $\chi_c$ shows a 5-fold enhancement at low temperatures, which is higher than in the Ni case. However, this is not the only difference between Co and other impurities investigated here. Perhaps, a more striking difference is related to the anisotropic behavior that emerges upon Co doping. 

\begin{figure}[!]
	\centering
	\includegraphics[width=0.5\textwidth]{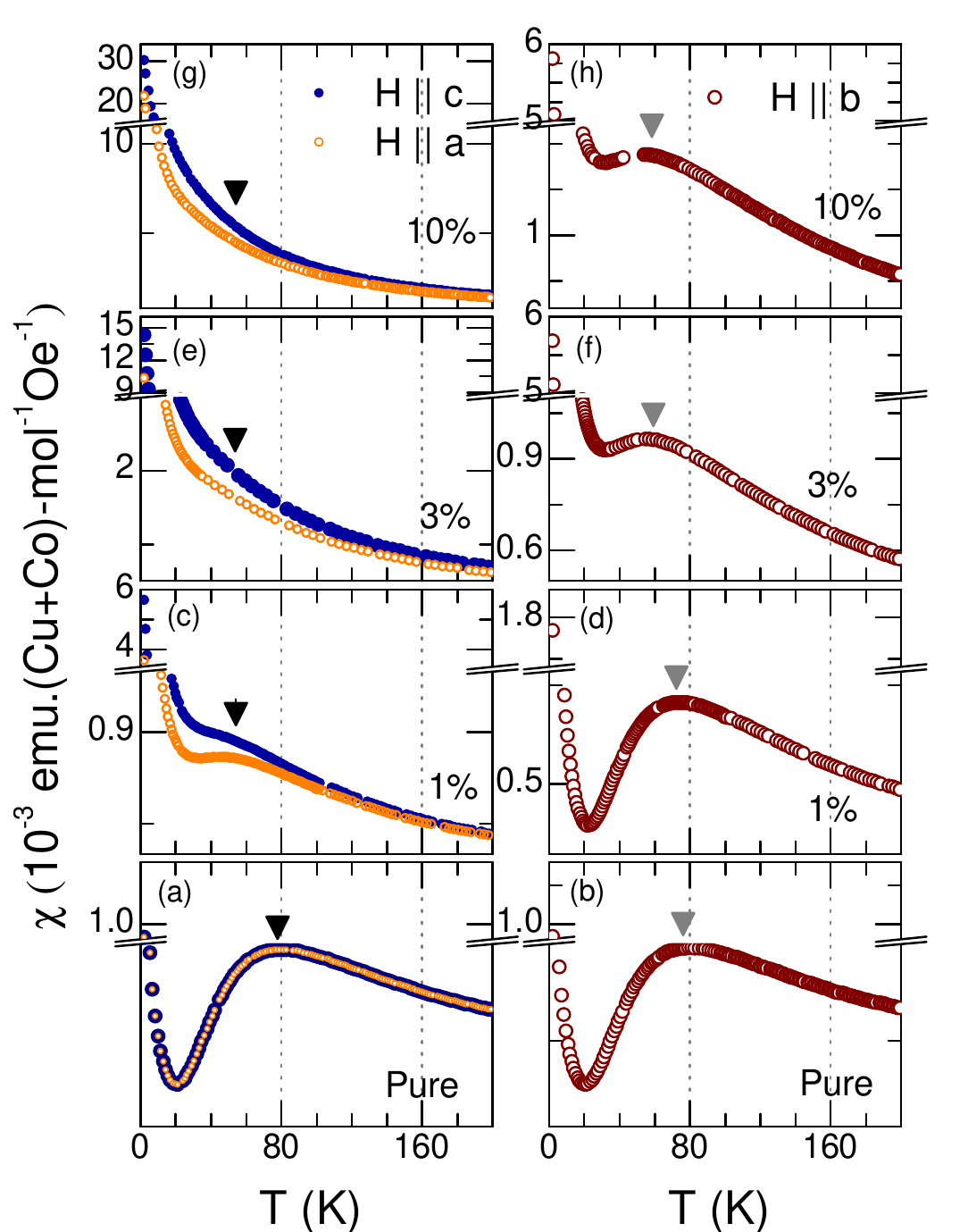}
	\caption{From (d) to (g) is shown the susceptibility ($\chi$) of Co doped Sr$_{14}$Cu$_{24}$O$_{41}$ single crystals along the crystallographic a, b and c-axis shown as a function of temperature. In (a) and (b) corresponding data for an undoped Sr$_{14}$Cu$_{24}$O$_{41}$ single crystal are also shown for comparison} 
	\label{chi_co}
\end{figure}

In Fig. \ref{chi_co}, we show $\chi_a$, $\chi_b$ and $\chi_c$ for 1, 3 and 10 \% Co doping. The susceptibility of pristine crystal is also shown for comparison. Since no qualitative change takes place between 3 to 10 \% doping range, for brevity we did not include the 5 \% data in Fig. \ref{chi_co}. We note that in the presence of Co impurity, the sign of magnetic anisotropy has reversed, i.e., $\chi_b < \chi_a \lesssim \chi_c$, which is opposite to the pristine case where $\chi_b > \chi_a = \chi_c$. 

For the 1\% sample, the ratio $\chi_c/ \chi_b$ is nearly 5 : 2. For higher Co-concentrations, the anisotropy is even more pronounced with $\chi_c/ \chi_b$ at T = 2 K approaching values as high as 5 : 1 (10\% Co). $\chi_a$ in each case shows a variation similar to that of $\chi_c$ but with a somewhat smaller magnitude.

Unlike a Cu$^{2+}$ ion in oxides and hydrides, which is characterized by a nearly isotropic spin 1/2, the unquenched residual orbital angular momentum in the Co ions can give rise to a significant single-ion anisotropy \cite{Slonczewski1961, Abragam1951}. In the spin chains or spin ladders, due to low-coordination, the anisotropy at the impurity site percolates via the spin-spin interaction along the chain length, causing the bulk of the crystal to show an anisotropic response. Recently, we found that merely 1 \% of Co doping in the spin chains SrCuO$_2$ and Sr$_2$CuO$_3$ induces a significant magnetic anisotropy in the bulk of the sample \cite{KarmakarPRL2017}.

Interestingly, in the doping range from 3\% to 10 \%, the value of $\chi_b$ at 2 K increases only marginally with x$_{Co}$ but over the same doping range the value of $\chi_c$ doubles. Another striking feature of the data shown in Fig. \ref{chi_co} concerns the position and size of the susceptibility maximum ($\chi_{max}$). While in the $\chi_b$ data, $\chi_{max}$ remains clearly discernible up to the highest Co doping (10\%), along the c-axis (or a-axis), above 3\% of Co-doping, $\chi_{max}$ is not easily discernible due to an overwhelming Curie-like background of the Co impurity spins.

\begin{figure}[!]
	\centering
	\includegraphics[width=0.43\textwidth]{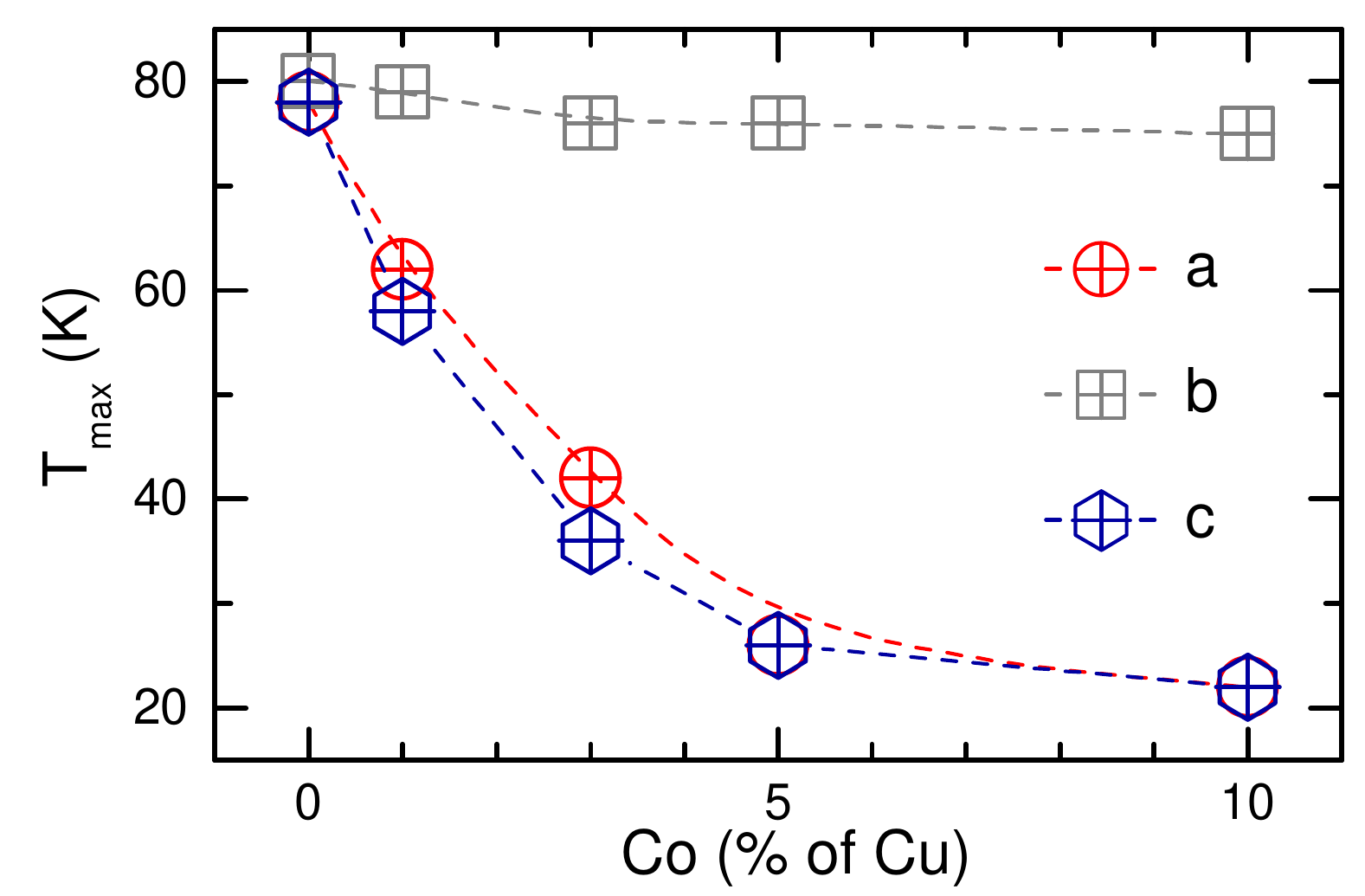}
	\caption{Variation of T$_{max}$ with Co concentration in the Co doped single crystals along the crystallographic a, b and c-axis (see text for details)} 
	\label{Co_J1}
\end{figure}

Trial fits using eq. \ref{2}, as done previously in the Zn and Ni case, did not give satisfactory results, particularly around $\chi_{max}$ in $\chi_c$ (and $\chi_a$). This probably suggests that Co doping, even at 1\% concentration, has a significant effect on the magnetic ground state of Sr$_{14}$Cu$_{24}$O$_{41}$. 

To make an approximate estimate of the dimer contribution to the total susceptibility, we subtracted the low-temperature Curie-tail from the measured $\chi$. This is done by fitting $\chi(T)$ in temperature range 2 K to 15 K using the Curie-Weiss law. From the subtracted data ($\Delta \chi$), shown in the supplementary information, T$_{max}$ is obtained.  

In Fig. \ref{Co_J1} we show the variation of T$_{max}$ as a function of x$_{Co}$ along the three crystal orientations. We find that along the b-axis, suppression of $T_{max}$ is smaller compared to that along the c (or a)-axis. Along these orientations T$_{max}$ decreases rather sharply, and by a significantly larger value. Interestingly, independent of the crystal orientation, T$_{max}$ tends to saturate beyond about 3\% of Co doping.    

Since the intradimer exchange, or the spin dimerization gap (J$_{ZR}$), is proportional to $T_{max}$ (J$_{ZR} \approx$ 5T$_{max}$/3 ), based on the variation of $T_{max}$ shown in Fig. \ref{Co_J1}, it can be inferred that the rate of suppression of J$_{ZR}$ in the Co-doped samples is rather anisotropic. With initial Co doping, the dimerization gap along the c-axis or a-axis closes at a much faster rate (27 \% per \% of Co doping) than along the b-axis (2-3\% per \% of Co doping). However, as the concentration of Co is increased beyond $\sim$3\%, $\Delta T_{max}/\Delta x_{Co}$ becomes very small, indicating a saturation. This observation suggests that for Co doping higher than 3\%, the Co impurity does not dope the chains, i.e., the concentration of Co in the chains saturate for x$_{Co}$ $>$ 3\%. We believe that at higher concentrations, Co is doped in the ladder sublattice.         

\subsubsection{Isothermal magnetization}
\label{MH1}

\begin{figure}[!]
	\centering
	\includegraphics[width=0.47\textwidth]{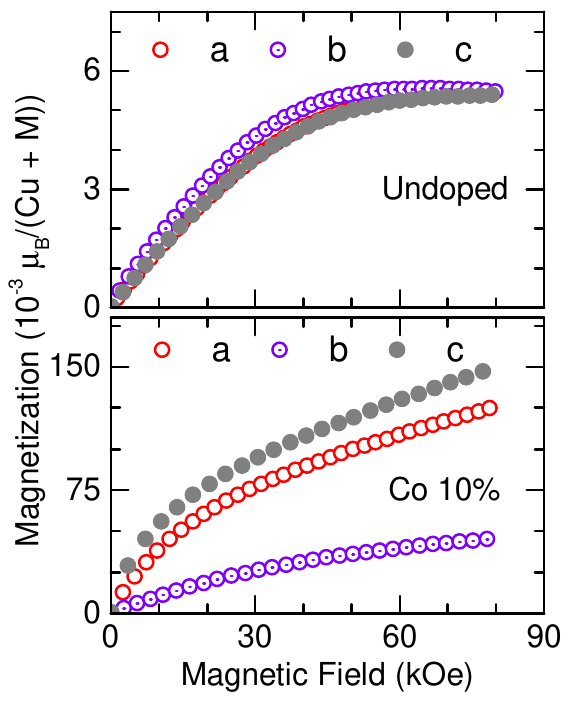}
	\caption{Isothermal magnetization M plotted as a function of applied magnetic field (H) at T = 2 K for an (a) undoped and (b) 10\% Co doped Sr$_{14}$Cu$_{24}$O$_{41}$ single crystals.} 
	\label{MH}
\end{figure}

The isothermal magnetization M$_\epsilon$(H) ($\epsilon$ = a, b and c) at T = 2K for a 10 \% Co doped crystal is shown in Fig. \ref{MH}. Data for other Co doped crystals are found to be similar and are shown in the supplementary section. For comparison, however, analogous data for the undoped crystal are also shown. In the undoped crystal, the magnetization behavior is isotropic in the ac-plane. A slight anisotropy with respect to the b-axis (M$_b$ $\gtrsim$ M$_a$ = M$_c$), is consistent with the $\chi$(T) data discussed earlier. 
 
In 10\% Co doped crystal, the magnetization in the ac-plane is different along the a- and c-axis. More importantly, the sign of magnetic anisotropy has reversed, such that, M$_b$ $<$ M$_a$ $\lesssim$ M$_c$, i.e., the b-axis is no longer the \textit{easy} magnetization axis. A similar behavior is found for 1, 3 and 5 \% crystals. 

\begin{figure}[!]
	\centering
	\includegraphics[width=0.52\textwidth]{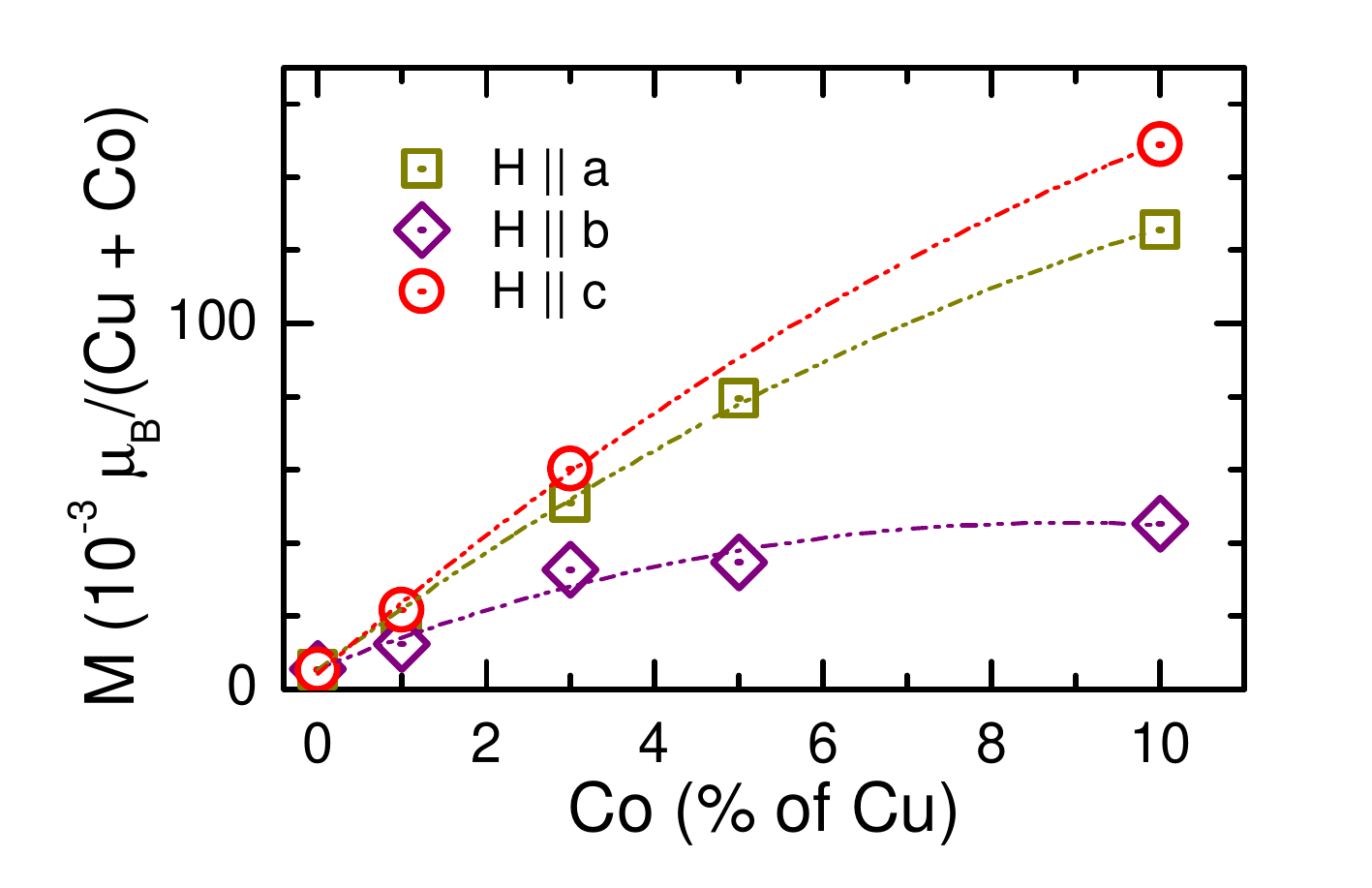}
	\caption{Magnetization at a field of 80 kOe for Co doped Sr$_{14}$Cu$_{24}$O$_{41}$ single crystals along a, b and c-axis plotted as a function of Co concentration.} 
	\label{MH_Co}
\end{figure}

In Fig. \ref{MH_Co}, we show the value of M(H)$|_{2K}$ at H = 80 kOe for 0, 1, 3, 5 and 10 \% Co doped crystals. We found that along the b-axis, the magnetization initially increases with Co-doping but levels-off at higher Co concentrations ($>$3\%). On the other hand, along the a- (or c-axis) it continues to grow almost linearly.

We recall that above 3\%, Co is not doped in the chains (inferred from the variation of T$_{max}$ in the previous section). Therefore, the observed saturation of magnetization along b-axis above 3\% doping suggests that the Co ions doped in the ladder have no spin component along the b-axis. On this basis one can hypothesize that Co spins in the ladder sublattice have a planar (ac-plane) anisotropy.    

At this point it is worth mentioning that the anisotropy for Zn, Ni and Al doped crystals has the same sign as for the undoped crystal, but it is opposite for the crystals doped with Co.           

\subsubsection{XPS}
\label{XPS}

While the magnetic behavior in the presence of Co impurity revealed useful information concerning the magnetic anisotropy induced by Co and its site-substitution preferences, the valence state of Co (i.e., whether it is in a +2 or a +3 valence state) still remains to be ascertained.

To ascertain the valence state of Co, we performed X-ray photoelectron spectroscopy (XPS) on the single crystal specimens of pristine and doped samples. From the survey scan it was confirmed that no other elements except Sr, Cu, Co, O and C are present. Calibration is done using the adventitious C 1S peak at 285.2 eV \cite{Wagner1979}. The Cu core data (not shown) consisted of two main peaks at 954.2 eV (Cu 2p$_{1/2}$) and 932.6 eV (Cu 2p$_{3/2}$), and a smaller and broader satellite peak at 944.7 eV (Cu 2p$_{3/2}$ satellite) in conformity with the standard data \cite{Calderon2008}. 

\begin{figure}[b]
	\centering
	\includegraphics[width=0.44\textwidth]{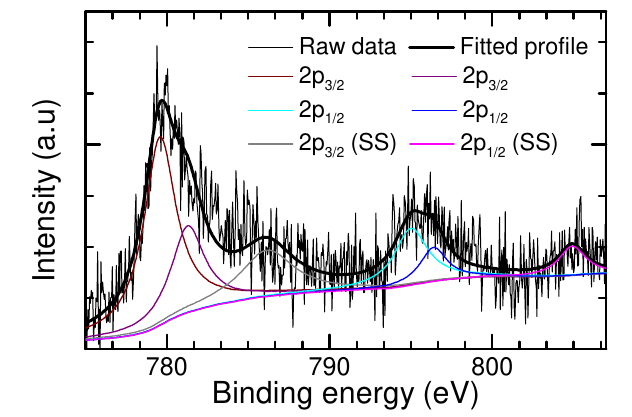}
	\caption{Co-2p core level spectrum showing 2p$_{3/2}$ and 2p$_{1/2}$ peaks and the shake-up satellite peaks for a 10\% Co-doped Sr$_{14}$Cu$_{24}$O$_{41}$ single crystal. The lines are fit to the data (see text for details).} 
	\label{xps}
\end{figure}

The XPS spectrum of a 10\% Co-doped crystal showing Co-2p peaks is presented in Fig. \ref{xps}. Due to low concentration of Co in the sample, the signal-to-noise ratio in our data is not particularly high. The crystals with lower Co-concentrations (not shown) suffered from even poorer signal-to-noise ratio and are therefore not considered further. Nonetheless, in the spectrum of 10\% Co-doped crystal in Fig. \ref{xps}, one can easily identify peaks at the binding energies close to 780 eV, 795 eV and 806 eV, with a broad and not so clearly discernible feature around 785 eV. 

Co-2p core level spectrum is split in 2p$_{3/2}$ and 2p$_{1/2}$ features due to spin-orbit coupling. The asymmetry in the peak shape towards the higher binding energy side in both, 2p$_{3/2}$ and 2p$_{1/2}$ features, indicates the possibility of mixed valance state. Therefore, the whole spectrum is fitted with \textit{six} Lorentzian peaks, and background is approximated by Shirley function. Comparing the binding energy positions with various valance states of Co as reported in available literature, we conclude that cobalt is present in two charge states: Co$^{2+}$ and Co$^{3+}$ (Refs. \citenum{TanJACS1991,Haber1977,Chuang1976}). The estimated ratio Co$^{3+}$/Co$^{2+}$ is approximately 3.5 : 1. The additional features labeled as SS correspond to the shake-up satellites of Co$^{2+}$ and Co$^{3+}$. Thus, in the crystal doped with 10 \% of Co, approximately 23\% of the doped Co ions are inferred to be in a +2 state and 77\% in +3 state. If we combine this with the result from a previous section that above approximately 0.003/Cu of doping concentration, Co is not doped in the chains, one can infer that in the 10\% Co doped crystal, the ratio Co$^{2+}$/Co$^{3+}$ in the chains is around 2.3 : 0.7.

\section{Summary and conclusions}
\label{sum}
We investigated the susceptibility of a pristine Sr$_{14}$Cu$_{24}$O$_{41}$ single crystal in considerable detail. We found that the "free" spins picture breaks down at low-temperatures, which is not surprising since these spins are weakly interacting. However, what is intriguing is that rather than ordering or freezing in some ordered pattern, the "free" spins prefer a spin dimerized state. Sahling et al.\cite{SahlingNature2015} proposed a quantum entangled ground state for the spin chains of Sr$_{14}$Cu$_{24}$O$_{41}$ wherein the “free” spins pair-up to form spin dimers over long-distances. We confirmed that this model indeed provides a satisfactory description of the magnetization of Sr$_{14}$Cu$_{24}$O$_{41}$ at low-temperatures. 

The spin chains in Sr$_{14}$Cu$_{24}$O$_{41}$, therefore, harbor two types of spin dimers: the dimers that form at high temperatures ($\sim$100 K) between two Cu$^{2+}$ spins separated by a ZR singlet ($\uparrow \bullet \downarrow$); and the dimers that appear at low-temperatures between a pair of "free" spins at large separations ($\uparrow \dots \downarrow$).

It is intriguing that the weak interaction between the "free" spins favors a dimerized state over an ordered or a quasi-ordered state. We speculate that this can be explained by considering that the "free" spins in the chain are randomly located, i.e., the distances, and hence J$_{LD}$, varies between the spin pairs. In such a scenario, when the sample is cooled down gradually to low temperatures, the spins that are closest form a spin dimer first. Upon further cooling, the spins at the next higher separation pair up, and so on and so forth. This process, in our view, is somewhat akin to the Random-singlet phase in a strongly disordered HAF spin 1/2 chain \cite{Kirkpatrick, Dasgupta}. 

We further examined how robust these dimers are against magnetic and non-magnetic impurities. For this purpose, we chose 4 different impurities, namely Zn, Ni, Al and Co.        

\textit{Zn impurity}: We found that 0.25 \% Zn doping has practically no effect on the Curie-tail and hence on the "free" spin density. This indicates that Zn dopes at the site of an interdimer ZR singlet ($\circ$). Doing so neither severs the dimers nor affects the "free" spins that were already present (due to hole transfer). Had Zn been doped at any other site in the chain, or in the ladders, new "free" spins would have been liberated, which would have enhanced the Curie tail contrary to the experimental result. In a previous paper\cite{LinPRB2001} on the polycrystalline samples, N$_d^{ZR}$ was reported to not change with small Zn doping, and our results are in agreement with it. 

For 1\% doping concentration the Curie-tail is found to be enhanced suggesting the breaking of ZRDs, which was confirmed by fitting the data above T = 10 K using the dimer model, which showed clear decrease in the value of N$_d^{ZR}$ and J$_{ZR}$. In another study\cite{WangPhysicaB} on the polycrystalline Zn doped samples, where the dimer model was used, $\sim$18\% decrease in N$_d^{ZR}$ is reported, which is comparable to our value ($\sim$11\%). However, in this study, the limit up to which Zn impurity does not sever the dimers in the chains was not studied.  
   
Substitution of a divalent Zn impurity for a ZR singlet $\circ$ in the chain (which can be naively thought of as Cu$^{3+}$) is expected to transfer a hole from the chains to ladders. And, it has been theoretically predicted that the excess holes in a ladder pair-up along the rungs\cite{Sigrist1994}. Therefore, the presence of holes in the ladder as paired entities does not contribute to the Curie-tail. This indeed is consistent with our experimental result on low level of Zn doping (Zn0.25 crystal). However, the presence of holes in the ladder sublattice should suppresses the magnetic thermal conductivity because hole-pair on the rung of a ladder will effectively cut the ladder into two decoupled segments, preventing the magnetic excitations from propagating through and hence diminishing the magnetic thermal conductivity. Indeed, it has been shown in a previous work \cite{HessPRB2006} that doping Sr$_{14}$Cu$_{24}$O$_{41}$ with Zn suppresses the magnetic thermal conductivity significantly.

\textit{Ni impurity}: The Ni impurity is found to be in a high-spin state. We examined its effect on the dimers $\uparrow \bullet \downarrow$ and $\uparrow \dots \downarrow$. We show that at 0.25 \% Ni doping, the parameters of ZRD term in eq. \ref{2} are nearly unchanged, suggesting that at low Ni concentration, Ni impurity has little or practically no effect on the dimer $\uparrow \bullet \downarrow$. The increase in the “free” spin density ($\Delta$N$_s$) is found to agree fairly nicely with the actual Ni concentration (x$_{Ni}$) in the crystal as determined using the ICP technique. From this observation we inferred that similar to the Zn case, at low concentrations, Ni impurity also substitutes the ZR singlets $\circ$ in the chains. 

At 1\% Ni doping, clear evidence of severing of the dimers $\uparrow \bullet \downarrow$ is found. Therefore, at this concentration, Ni substitutes not only $\circ$ but also $\uparrow$ (or $\downarrow$) and/or $\bullet$ which breaks the dimers liberating "free" spins. The value of N$_d^{ZR}$ is found to decrease by about 9\% and that of J$_{ZR}$ by about 3 to 4\%. In a previous polycrystalline work \cite{WangPhysicaB}, N$_d^{ZR}$ was reported to decrease by almost 28\%, and J$_{ZR}$ to increase to about 10\%. Both these values appear mutually inconsistent because a 28\% decrease in N$_d^{ZR}$ due to severing of dimers cannot simultaneously lead to an increases in the intradimer exchange. Due to dimer breaking the average value of J$_{ZR}$ is expected to decrease as is found to be the case in our study.
      
The specific heat data down to T = 0.06 K on the 0.25\% Ni doped crystal is compared with that of the pristine Sr$_{14}$Cu$_{24}$O$_{41}$ crystal. We found that in the Ni doped crystal, the broad specific heat anomaly associated with the long-distance dimers is suppressed by about 25\%. This is quite interesting since at a doping concentration of 0.25\%, the average separation between the Ni impurities in the chain is about 400 Cu sites, which is rather large. However, given that the average size of the dimers $\uparrow \dots \downarrow$ is about 100 Cu sites ($\sim$200 \AA)\cite{SahlingNature2015}), there is a reasonably high probability of finding a Ni impurity intercepting one out of four randomly picked $\uparrow \dots \downarrow$ dimers in the chain. This suggests that the long-distance dimers are severed in the presence of Ni impurity; however, the dimers without a Ni impurity enclosed between the pairing spins remain intact. From this one can infer that the effect of Ni impurity is highly local as it severs only the $\uparrow \dots \downarrow$ that it occupy; and hence the  presence of Ni impurity does not completely disrupt the quantum entangled ground state of the spins chains. How exactly the presence of Ni spin severs a dimer is something that should be pursued further.         

\textit{Al impurity}: The case of trivalent Al$^{3+}$ impurity turned out to be different from the divalent impurities Zn and Ni. We found that unlike Zn or Ni, the solid-solubility of Al in the crystal is very limited. The crystal with 1\% nominal Al concentration is found to contain about 0.25\% of Al. However, what makes it different from the divalent Zn and Ni impurities is that Al prefers to dope the ladders rather than the chains. This is inferred from the fact that while the Curie-tail is enhanced upon Al doping, the values of N$_d^{ZR}$ and J$_{ZR}$ did not change. Al impurity in the ladder should break a spin-singlet along the rung of the ladder, releasing free spins. This will enhance the Curie-tail without altering the dimers in the chains, as is found to be the case experimentally.      

\textit{Co impurity}: The strongest effect of doping on the magnetic behavior is observed in the presence of the Co impurity. Co-doping not only enhances the low-temperature Curie-tail but also results in a highly anisotropic magnetic behavior. We argued that this strong anisotropy, even for doping level as low as 1 \%, is due to the single-ion anisotropy of the Co ion. We also found evidence to suggest that upon increasing the Co concentration beyond to $\sim$3\% , the population of Co in the chain sublattice reaches a saturation level. Interestingly, the dimerization gap associated with the Zhang-Rice dimers also turned out to be highly anisotropic with a much higher suppression rate in the ac-plane than along the b-axis. 

The behavior of M$_{2K}$(80 kOe) or that of T$_{max}$ with Co-concentration, corroborated by the XPS data, suggest the presence of two types of Co ions: Co$^{3+}$ in the ladders having a planar (ac-plane) anisotropy, and Co$^{2+}$ in the chains with a spin component along the b-axis. This is also in agreement with the result that trivalent Al prefers to dope the ladders and divalent Ni and Zn the chains. 

In summary, we investigated the effect of impurities on the magnetic behavior of the hybrid chain/ladder compound Sr$_{14}$Cu$_{24}$O$_{41}$. We observed profound changes in the magnetic behavior in the presence of an impurity. The exact manner in which the magnetic behavior is altered in the presence of an impurity varied from one impurity to another.

The effect of Ni impurity on the long-distance dimers $\uparrow \dots \downarrow$ is investigated using the specific heat probe down to T = 0.06 K. We found that the presence of Ni impurity in the dimers $\uparrow \dots \downarrow$ severs them; but the dimers without a Ni impurity remain intact. Hence we conclude that while the dimers themselves are are quite sensitive to the presence of a Ni impurity between the pairing spins, the quantum entangled ground state of the spin chains in Sr$_{14}$Cu$_{24}$O$_{41}$ is robust and is not completely disrupted even at 0.25\% of Ni doping. On the other hand, at the lower impurity concentration the high-temperature dimers $\uparrow \bullet \downarrow$ are found to be practically unaffected; but when the impurity concentration is raised to 1\% these too were found to be significantly severed.            

Doping with magnetic Co impurity in Sr$_{14}$Cu$_{24}$O$_{41}$ resulted in reversing the sign of the magnetic anisotropy. The dimerization gap (J$^b_{ZR}$/J$^c_{ZR}$) associated with the Zhang-Rice dimers is found to exhibit an anisotropic variation with increasing Co doping, reaching a maximum value  of approximately 1 : 4, as shown in Fig. \ref{Co_J1}, between 3\% to 5\% doping. At higher Co concentrations, the dimerization gap does not change, suggesting that the upper limit of Co concentration in the chain sublattice is close to 3\%.      
	
		\bibliographystyle{apsrev4-1}
		\bibliography{RB_PRB}
	
\end{document}